# On the origin of magnetism in (Ga,Mn)As: from paramagnetic through superparamagnetic to ferromagnetic phase

L. Gluba[1,2], O. Yastrubchak[3,2*], J.Z. Domagala[4], R. Jakiela[4], T. Andrearczyk[4], J. Żuk[2], T. Wosinski[4], J. Sadowski[5,4,6] and M. Sawicki[4]

[1]*Institute of Agrophysics, Polish Academy of Sciences, Doświadczalna 4, 20-290 Lublin, Poland*

[2]*Institute of Physics, Maria Curie-Sklodowska University in Lublin, Pl. M. Curie-Skłodowskiej 1, 20-031 Lublin, Poland*

[3]*V.E. Lashkaryov Institute of Semiconductor Physics, National Academy of Sciences of Ukraine, 03028 Kyiv, Ukraine*

[4]*Institute of Physics, Polish Academy of Sciences, Aleja Lotnikow 32/46, PL-02668 Warsaw, Poland*

[5]*MAX-IV laboratory, Lund University, P.O. Box.118, 22100 Lund, Sweden*

[6]*Department of Physics and Electrical Engineering, Linnaeus University, SE-391 82 Kalmar, Sweden*

The high-spectral-resolution spectroscopic studies of the energy gap evolution, supplemented with electronic, magnetic and structural characterization, show that the modification of the GaAs valence band caused by Mn incorporation occurs already for a very low Mn content, much lower than that required to support ferromagnetic spin – spin coupling in (Ga,Mn)As. Only for *n*-type (Ga,Mn)As with the Mn content below about 0.3% the Mn-related extended states are visible as a feature detached from the valence-band edge and partly occupied with electrons. The combined magnetic and low-temperature photoreflectance studies presented here indicate that the paramagnetic - ferromagnetic transformation in *p*-type (Ga,Mn)As takes place without imposing changes of the unitary character of the valence band with the Fermi level located therein. The whole process is rooted in the nanoscale fluctuations of the local (hole) density of states and the formation of a superparamagnetic-like state. The Fermi level in (Ga,Mn)As is coarsened by the carrier concentration of the itinerant valence band holes and further fine-tuned by the many-body interactions.

**Introduction**

(Ga,Mn)As dilute ferromagnetic semiconductor (DFS) has been a subject of excessive interest for the last two decades, because of its unique properties and relatively high Curie temperature ($T_C$) close to 200 K.[1] The ability to control the magnetic state through electric field gating,[1–5] light exposure[6–8] or pressure,[9,10] brought (Ga,Mn)As to a prototype material for semiconductor spintronics.[11] Nevertheless, despite proved spintronic capabilities revealed by many diverse experiments, the ultimate view of the band structure and the origin of free holes in this material is still being debated.

---

* Corresponding author: Oksana Yastrubchak; E-mail: plazmonoki@gmail.com

Two main models of the band structure of zinc blende (Ga,Mn)As have been proposed. The first one, the kinetic p-d Zener model, assumes merging the Mn impurity states with the GaAs host valence band (VB) and the Fermi level located in the VB settled by the concentration of free holes.[12,13] The other one, the impurity band (IB) model, assumes formation of the Mn-related IB above the GaAs VB edge with the Fermi level pinned within the IB. In the latter model, the ferromagnetic (FM) interactions can be explained by the double-exchange mechanism, which assumes hopping conduction involving the IB holes. In addition, there is no agreement whether the IB is detached[14–16] or merged[17–24] with the VB states. Unfortunately, the picture of the Mn content driven band structure crossover from insulating to metallic state at the onset of ferromagnetism, has not been fully understood over the last two decades, despite much efforts. Thus a better understanding of the band structure at low Mndoping level is highly desirable.

Due to the complicated nature of this compound it is difficult to decisively point out which from the above two mentioned mechanisms is responsible for the magnetic properties of (Ga,Mn)As. Recently, the DFS community focused efforts on direct probing of the VB states.[15,17–24] In this paper we present the results of investigations of the optical-gap behaviour in (Ga,Mn)As in conjunction with the evolution of structural, electrical and magnetic properties due to increasing Mn doping level.

**Experimental**

To capture the Mn-induced modifications of the fundamental properties of (Ga,Mn)As, it is essential to understand the nature of Mn states in GaAs. To this end a set of (Ga,Mn)As layers has been prepared by the low-temperature molecular-beam epitaxy (MBE) growth technique, with the Mn contents ($x$) ranging from 0 (an LT-GaAs reference layer) to 1.6%. The direct information on the influence of Mn doping on the (Ga,Mn)As band structure can be obtained by the analysis of the optical gap, which have been measured using low-temperature photoreflectance (LT-PR).[25] This nonlinear optical technique gives sharp, derivative-like spectral features. It provides an improvement in the accuracy over our earlier photoreflectance studies performed at room temperature.[26–29] LT-PR offers also an advantage over other optical methods applied to (Ga,Mn)As[30–34] where the optical response from the gap may be quenched by defects and is marred by the hole plasma. Importantly, the method allows to probe the valence-to-conduction band optical transitions regardless of the Fermi level location within the gap. Moreover, the analysis of the optical transition energies enables elucidation of the behaviour of the Fermi level in *p*-type samples depending on the acceptor-doping level. The LT-PR results are supported by magnetization measurements using superconducting quantum interference device (SQUID) magnetometry and the Raman scattering experiments (sensitive to hole density), in order to reveal the correlation between magnetic, electronic and optical properties of (Ga,Mn)As over its evolution from paramagnetic (PM) through superparamagnetic (SP) to ferromagnetic phase with increasing Mn content. Finally, in order to examine the VB dispersion and the density of states in a low-Mn-doped ferromagnetic layer *in-situ* angle-resolved photoemission spectroscopy (ARPES) measurements have been performed. The Mn content in all the (Ga,Mn)As layers has been determined by secondary ion mass spectrometry (SIMS) and high-resolution X-ray diffraction (HR-XRD).

**Results**

High-resolution X-ray diffraction measurements show that investigated LT-GaAs and (Ga,Mn)As epitaxial layers have been pseudomorphically grown on GaAs substrate. Due to the Mn-induced increase in the (Ga,Mn)As lattice parameter[35] they are under compressive misfit strain. As typical for (Ga,Mn)As and LT-GaAs epilayers deposited on GaAs substrates, the layers are fully/coherently strained, i.e. they have the same in plane lattice parameter as that of the substrate. Fig. 1 demonstrates 2θ/ω curves for the 004 Bragg reflections for all investigated samples. Clear X-ray interference fringes visible for the (Ga,Mn)As layers with Mn content $x \geq 0.9\%$ imply a high structural perfection of the layers and good quality of the interfaces. The layer thickness calculated from the angular spacing of the fringes for the $x = 1.6\%$ layer corresponds very well to the value derived from the MBE growth parameters and SIMS measurements. On the other hand, the lack of X-ray fringes for the layers with $x \leq 0.3\%$ (including the LT-GaAs) indicates a very small difference between the lattice parameters of these layers and the GaAs substrate, highlighting a rather low concentration of the arsenic antisite, $As_{Ga}$, defects - in the range of $10^{19}$ cm$^{-3}$.[35] Indeed the MBE growth was thoroughly optimized in order to minimize the concentration of $As_{Ga}$ in LT-GaAs and (Ga,Mn)As. The samples were grown in close-to-stoichiometric conditions, i.e., with carefully calibrated (Mn + Ga) vs As flux ratios; entirely based on RHEED intensity oscillations measured both for test samples and during the growth of actual LT-GaAs and (Ga,Mn)As layers.

Diffraction peaks corresponding to (Ga,Mn)As epitaxial layers shift to smaller angles, with respect to that of the GaAs substrate, as a result of larger perpendicular lattice parameters. Their angular positions have been used to calculate the perpendicular ($a_\perp$) and the relaxed lattice parameters ($a_{rel}$) of the layers, assuming the same elasticity constants of (Ga,Mn)As as of GaAs.[36,37] In addition, the out-of-plane strain in the layers, defined as $(a_\perp - a_{rel})/a_{rel}$, have been calculated and listed in Table 1. This table summarizes also the values of Mn contents estimated by HR-XRD and SIMS. We note quite a good correspondence between the two methods. Throughout the rest of the paper the $x$ values obtained by SIMS are used to label the layers.

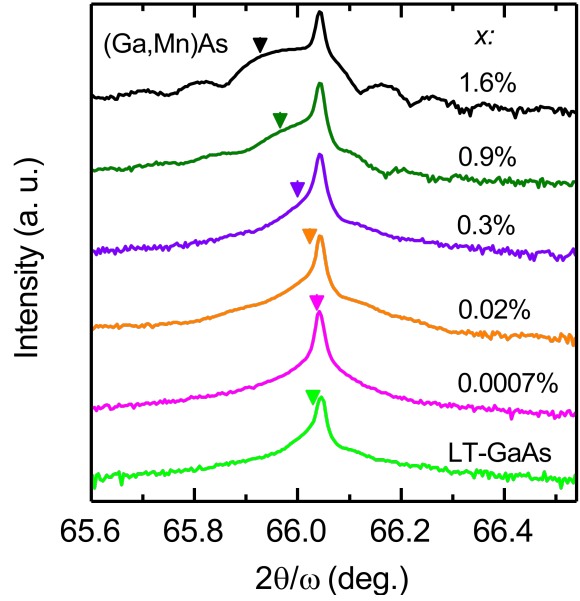

**Figure 1.** High-resolution X-ray diffraction patterns of the studied layers: 004 Bragg reflections, 2θ/ω scans ($\lambda_{Cu\ K\alpha 1}$) for the LT-GaAs and (Ga,Mn)As layers. The central narrow features correspond to reflections from the GaAs substrate and the broader peaks at lower angles, marked with arrows, are reflections from the layers. Their positions for the LT-GaAs and (Ga,Mn)As layers with $x = 0.0007\%$ and $0.02\%$ are nearly indistinguishable from those of the GaAs substrate. Mn contents quoted in the figure have been obtained from the secondary ion mass spectrometry (SIMS) results.

The results of the micro-Raman scattering spectroscopy are shown in Fig. 2. In *p*-type GaAs, as well as in (Ga,Mn)As, longitudinal-optical (LO) phonon mode couples with the hole-gas-related plasmon forming so-called coupled plasmon–LO phonon mode (CPPM).[38] However, we do not observe this feature in the spectra of the LT-GaAs and (Ga,Mn)As layers with $x = 0.0007\%$ and $0.02\%$ - here only a strong LO phonon line and a very weak transverse-optical (TO) phonon line located around 290 cm$^{-1}$ and 265 cm$^{-1}$, respectively, is seen. It is well known that an excess of arsenic - mainly in the form of arsenic antisites, As$_{Ga}$,[39] builds into GaAs at low temperature MBE growth conditions. As$_{Ga}$ act as deep double donors and lead to an *n*-type hopping conductivity in LT-GaAs. Similarly, very diluted (Ga,Mn)As is *n*-type too, as shown for a comparable set of samples characterized by thermoelectric power measurements.[28] A threshold Mn concentration bordering *n* and *p*-type materials depends on the actual concentration of As$_{Ga}$ donors and so on the actual MBE chamber set up and the growth conditions. In the case of the present set of samples about 0.3% of Mn was needed to form the CPPM band in the Raman spectra, indicating that around this concentration of Mn (Ga,Mn)As turns *p*-type. A similar crossover from a fully compensated to the *p*-type (Ga,Mn)As for $x \cong 0.3\%$ was observed earlier as an anomalous behaviour of the (Ga,Mn)As lattice parameter dependence on the Mn content.[35] With further increase in *x* the CPPM mode starts to dominate the spectra, simultaneously with its energy shifted towards the TO-phonon-line wavenumber. This is the direct indication of an increasing holes density with *x*, which can be quantified from the full line-shape fitting. The procedure has been performed using generalization of Drude theory of the CPPM, assuming LO-phonon damping, effective masses of light and heavy holes for GaAs and plasmon damping.[40,41] The values of plasmon damping have been estimated basing on the typical carrier mobilities i.e. 15 cm$^2$/Vs (*x* = 0.3% and 0.9%) and 9 cm$^2$/Vs for sample with *x* = 1.6%.[42] The hole densities determined this way are given in Fig. 2 and listed in Tab. 1.

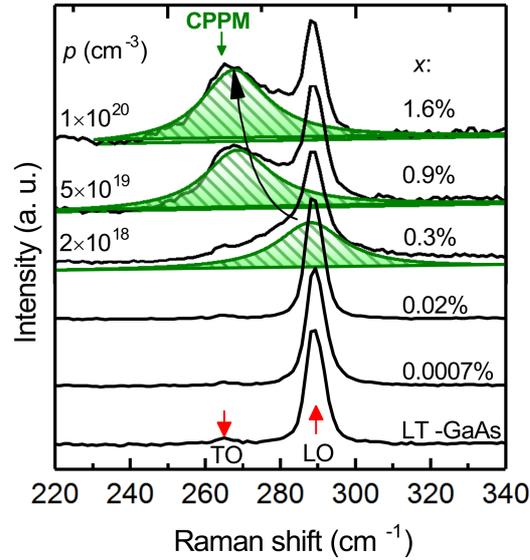

**Figure 2.** Micro-Raman spectra of the studied LT-GaAs and (Ga,Mn)As epitaxial layers. Apart from the strong longitudinal optical (LO) and transverse optical (TO) (weak) phonon modes, a coupled plasmon–LO phonon mode (CPPM – hatched peaks) is seen, which indicates a *p*-type character of the layers with *x* ≥ 0.3%. The estimated hole densities (in cm$^{-3}$) are given on the left hand side of the corresponding spectra.

**Table 1.** Mn contents obtained from the SIMS and HR-XRD results, values of the out-of-plane strain obtained from the HR-XRD results and hole densities estimated from the Raman spectra for LT-GaAs and (Ga,Mn)As layers.

| % Mn (SIMS) | % Mn (HR-XRD) | Out-of-plane strain ± 0.1 (×10$^{-4}$) | Hole density (cm$^{-3}$) |
|---|---|---|---|
| 0 (LT-GaAs) | - | 0.2 | - |
| 0.0007 | - | 0.2 | - |
| 0.02 | - | 0.9 | - |
| 0.3 | 0.3 | 1.4 | 2×10$^{18}$ |
| 0.9 | 0.7 | 3.2 | 5×10$^{19}$ |
| 1.6 | 1.4 | 6.2 | 1×10$^{20}$ |

Despite our best efforts[43] SQUID magnetometry has not yielded any indication for ferromagnetic spin-spin coupling down to 1.8 K in the studied layer with $x$ up to 0.3%. Specifically, neither open hysteresis curves nor a nonzero remnant moment have been observed, down to the ultimate magnetometer sensitivity of $5 \times 10^{-9}$ emu. This indicates that up to this doping level the Mn subsystem retains predominantly PM properties. The possible short range superexchange antiferromagnetic $Mn^{2+}$ - $Mn^{2+}$ or ferromagnetic $Mn^{3+}$ - $Mn^{3+}$ couplings evaded detection due to an insufficient amount of Mn species present for such a low doping levels in 100 nm thin layers.[44] Interestingly, only the PM is seen in $p$-type 0.3% sample, however, we note here that in the studied set of samples, the 0.3% one has only "just" turned $p$-type, and that such a low hole concentration, of the order of $2 \times 10^{18}$ $cm^{-3}$ (well below the Mott critical concentration, $p_C \sim 10^{20}$ $cm^{-3}$)[1] is insufficient to bring up noticeable long-range spin-spin interactions at the studied temperature range.

Remarkably, a further three-fold increase in $x$ (up to 0.9%) and in $p$ up to $p_C$ brings in a SP-like response. Its presence is revealed by the weak field temperature-dependent studies summarized in Fig. 3.

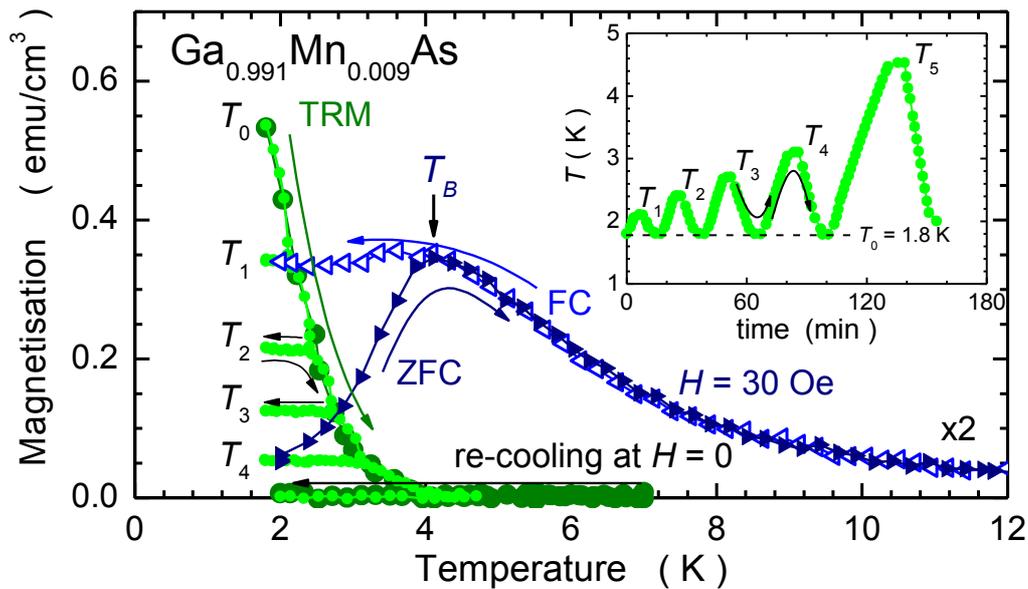

**Figure 3.** Low field temperature dependent studies of $x = 0.9\%$ (Ga,Mn)As layer. Circles indicate thermoremnant magnetisation (TRM) measured in two modes. Dark ones are obtained during a single shot warming from the base temperature ($T_0 = 1.8$ K) until above the signal vanishes and during re-cooling to $T_0$ at the same zero field conditions. Light circles indicate thermal cycling of the TRM between $T_0$ and progressively higher temperature ($T_1$ through $T_4$) according to the pattern presented in the inset. The triangles indicate temperature dependences of magnetisation measured at 30 Oe according to the well-established routine of a zero field cooled (ZFC – full triangles) and field cooled (FC – open triangles) states of the sample. $T_B$ marks the mean blocking temperature and the arrows indicate the directions of temperature sweeps.

The nonzero at the base temperature and quickly decaying with temperature remnant signal (the thermoremnant magnetisation, TRM) associated with a clear maximum present on the zero-field cooled (ZFC) trace are the main features in this figure.

Such behaviour strongly suggests a magnetically composite (granular) constitution of the sample. The (blocked) SP characteristics indicate that the FM coupling is already present in this sample but is maintained only in mesoscopic volumes which are dispersed in otherwise paramagnetic host. The blocking feature emerges either due to the magnetic anisotropy of these volumes (as discussed later) or due to their mutual dipolar interaction. The average size of these magnetic entities can be assessed from the magnitude of the temperature at which the maximum on the ZFC is seen (4 K). This so-called (mean) blocking temperature ($T_B$) of the distribution is related to the (mean) volume ($V$) in which this local FM coupling is maintained through: $25k_BT_B = KV$, where $K$, the anisotropy constant in (Ga,Mn)As ranges between 5000 to 50000 erg/cm$^3$,[45] $k_B$ is the Boltzmann constant, and the factor 25 is set by the experimental time scale – about 100 s, in SQUID magnetometry. From these numbers, we conclude that this mean volume corresponds to a sphere of a diameter between 8 to 20 nm, what indeed confirms a mesoscopic extent of the FM coupling. Further evidence confirming the magnetically inhomogeneous structure of this layer comes from thermal cycling of the TRM (detailed in the Methods, and presented in Fig. 3). This experimental protocol reveals that the remnant magnetisation gets visibly reduced only during the incremental warmings (from $T_i$ to $T_{i+1}$), following the original TRM. Otherwise, it stays fairly constant, being in particular insensitive to $T$-variations in any directions between the current $T_i$ and $T_0$. This clearly indicates the decisive role of the thermal agitation over the individual energy barriers specific to each mesoscopic FM volume. Finally, a lack of any magnetic moment exerted by this sample during re-cooling at the same $H = 0$ from above a temperature at which the TRM dropped to zero confirms the absence of a long range magnetic order, despite the existence of a non-zero TRM at the first place. Interestingly, the magnetic properties of the $x = 0.9\%$ sample correspond qualitatively to the properties of (Ga,Mn)As shells overgrown on (In,Ga)As nanowires.[46] The common denominator for these two morphologically different (Ga,Mn)As systems is their insufficient hole density to globally support the long range FM order, however, sufficiently large to maintain it locally, on mesoscopic length scales.[47]

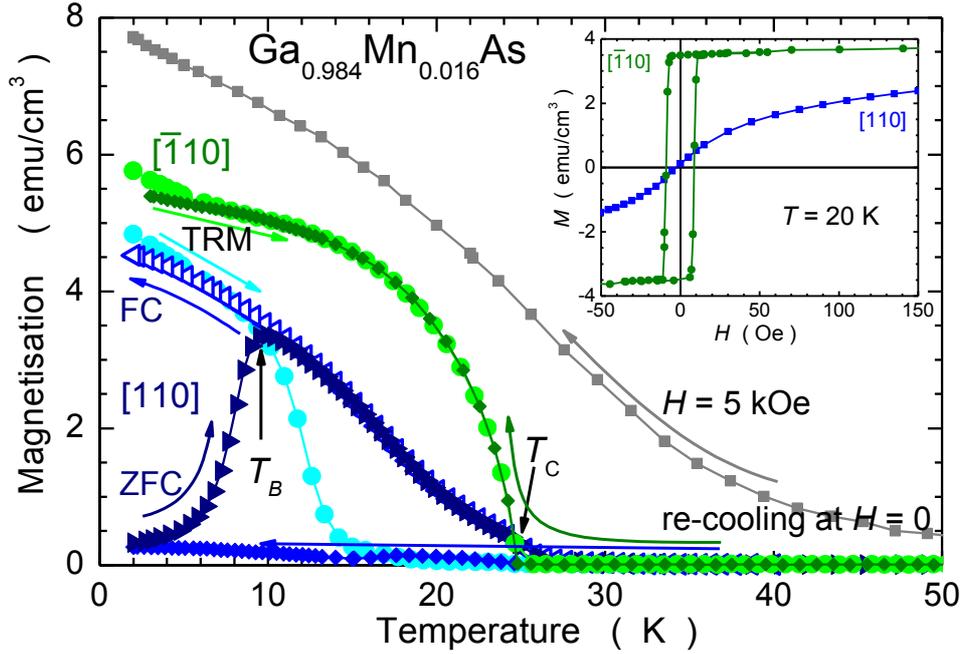

**Figure 4.** A coexistence of mesoscopic and microscopic ferromagnetic orders in $x = 1.6\%$ (Ga,Mn)As layer. Bullets and diamonds of light and dark hues of green and blue indicate thermoremnant magnetisation (TRM) and the re-cooling measurement to the base temperature at the same $H = 0$ condition along [-110] and [110] in-plane orientations, respectively. $T_C$ marks the Curie temperature for this sample. The triangles indicate temperature dependences of magnetisation measured at 30 Oe according to the well-established routine of zero field cooled (ZFC – full triangles) and field cooled (FC – open triangles) states of the sample. $T_B$ marks the mean blocking temperature. Full squares indicate magnetization at 5 kOe. The arrows indicate the directions of temperature sweeps. The inset shows isothermal magnetization field sweeps at 20 K for [-110] (bullets) and [110] (squares).

The next incremental step up in $x$ to 1.6% brings in a new feature, the global extent of the FM coupling. We substantiate this claim by observing that for this sample, as exemplified in Fig. 4, the TRM measurement performed along [-110] in-plane direction shows now a Brillouin-like concave curvature and a quite rapid roll-off to zero at a certain temperature, which we identify as the $T_C$ for this system. More importantly, on the re-cooling from above $T_C$ the data follow exactly the set recorded during the initial TRM. Such an observation of the appearance of a spontaneous magnetization ($M_S$) at zero field conditions below a certain temperature is the direct prove of the existence of the global extent of the FM order. It should be noted, however that the same magnitude of both remnant and spontaneous magnetisations is only possible due to the existence of a relatively strong in-plane uniaxial magnetic anisotropy (UMA)[45,48] in (Ga,Mn)As, as exemplified in the inset to Fig. 4. We finally infer from Fig. 4 that despite the presence of the strong UMA the magnitude of both $M_S$ and TRM is markedly smaller than the saturation magnetisation, approximated here as the magnetization recorded at $H = 5$ kOe - meaning that only a part of the Mn spins present in the layer contributes to this global coupling.

In order to establish the magnetic properties of these spins which do not contribute to $M_S$, the temperature dependent studies at $H = 0$ along the uniaxial hard, the [110] direction, have been performed (guided by the ref. 2), as owing to the UMA, no magnetic response specific to $M_S$ is expected at these conditions. On the contrary, as documented in Fig. 4, a magnetic remanence, comparable in magnitude to that recorded along the easy axis, is seen at the lowest temperatures. However the presently observed TRM vanishes substantially quicker and on re-cooling virtually none magnetic signal has been picked up by the magnetometer. We finally note that, qualitatively, the magnetic response along [110] direction is the same as that of the $x = 0.9\%$, the SP sample.

To further confirm the SP-like nature of this additional magnetic signal in the sample in which otherwise a clear thermodynamic phase transition is seen, analogously to the previous sample, ZFC – FC temperature cycling has been performed at 30 Oe. These measurements, when performed for [-110] orientation (for which the $M_S$ is probed) gave practically the same results as those obtained during $H = 0$ temperature cycling (the TRM and the subsequent re-cooling, not shown in Fig. 4 for the clarity of presentation) – an expected results for a magnetic system at the saturation. However, the ZFC - FC measurement performed along the orthogonal, [110] - the FM hard, direction yielded a strong response exhibiting all features seen already in the SP sample. Namely a clear maximum on ZFC related to dynamical blocking and a strong bifurcation between ZFC and FC developing below $T_B = 10$ K. This higher value, corresponding to somewhat larger diameter of 10 to 25 nm, is understood to be related to the lager $x$ and $p$ in this layer.

Two important facts should be stressed here. Firstly, we note that this sample directly corresponds to the (Ga,Mn)As sample studied by gating in ref. 2, where it was shown that despite a homogenous Mn distribution - for hole densities corresponding to the localization boundary both global and mesoscopic FM orders coexist in one sample and that relative contribution of these FM and SP components is primarily hole concentration dependent. However, these two samples were brought into the verge of localization by two different ways. The sample from ref. 2 is a high $x$ and high $p$ very thin (Ga,Mn)As layer in which localization threshold was induced by a relatively sizable surface depletion caused by the native oxide (consuming actually nearly top 50% of the whole 3.5 nm thin layer). In the 100 nm thick layer studied here the localization threshold is imposed (globally) by a low magnitude of $p$ assured by a rather low magnitude of $x$, and, likely, further reduced by compensating point defects such as $As_{Ga}$. Secondly, the possibility of such a clear cut experimental separation between the SP and the FM components coexisting in one material is possible only due to mutually orthogonal magnetic easy axes of both components, a feature brought about by a strong hole density dependent magnitude and sign of the UMA in (Ga,Mn)As. It involves two spin reorientation transitions (SRT), one of them taking place for low $p$ values,[49,50] We argue therefore that it is that low $p$ SRT of the UMA which differentiates the spatial orientation of the magnetic easy axes of the mesoscopically limited FM order (of overall SP characteristics) developing in regions of the sample with reduced hole densities[47] from the global FM one brought about in that part of the sample in which itinerant holes dwell. In (Ga,Mn)As this situation takes place for $p < 6 \times 10^{20}$ cm$^{-3}$, since above this hole density the uniaxial magnetic easy axis of the long range part rotates again to [110]).[49]

To conclude this part we note that upon further increase in $x$ and hence $p$, augmented sizably towards the limit of $p = xN_0$ by the low temperature annealing[51,52] the SP-like contribution leads completely to the global itinerant ferromagnetism whose micromagnetic properties are now very well understood.[45,53]

Photoreflectance (PR) studies have been performed in order to resolve the optical-energy-gap evolution in (Ga,Mn)As with increasing Mn content modifying also their magnetic and electronic properties. To obtain high energy resolution of PR spectra we have performed systematic studies at low temperatures (10 K) (Fig. 5). Performing PR measurements at low-temperatures, enables getting rid of the Franz-Keldysh oscillations - usually occurring at room temperature, and a sizable reduction of the thermal broadening. Hence with LT-PR one obtains a significantly higher energy resolution around the band gap, a merit unavailable by earlier spectroscopic measurements.[30–34] Additional advantage is that the LT-PR directly probes (Ga,Mn)As in the ferromagnetic phase - the case of $x = 1.6\%$ layer.

The collected raw derivative-like LT-PR modulated reflectivity ($\Delta R/R$) spectra, presented by open circles in Fig. 5a have been fitted using Aspnes's third derivative line-shapes (TDLS).[54,55] The deconvolution procedure has been performed using the lowest amount of TDLS features (max. 3) that maximizes the coefficient of determination $r^2$ (deconvolution shown in Supplementary Information, Fig. S1). Measurements probed with an approximately 0.0003 eV step ensure unique fits for the three fitted TDLS features with 4 free parameters for each one. The single spectral components have been integrated using the Kramers-Kronig (KK) relations.[56] This procedure facilitates an analysis of the critical-point energies by tracking a maximum of the KK modulus, shown in Fig. 5b. As expected, the LT-PR spectrum of the reference LT-GaAs layer exhibits two contributions (marked with diagonal hatching in Fig. 5b of similar energy, representing optical transitions between the heavy (HH) and light-hole (LH) valence sub-bands and the conduction band (CB). Relative intensities of the peaks imply a higher density of states of the HH sub-band than in LH sub-band. In turn, the LT-PR spectra of (Ga,Mn)As with $x = 0.0007\%$ and $0.02\%$ display, in addition to the two valence-band contributions, additional lower-energy peak (denoted by the dark solid filling in Fig. 5b. Moreover, the valence-band-related optical transitions differ from those of the LT-GaAs spectrum – they are energetically separated and blue-shifted with respect to those of the undoped layer. Additionally, the widths of the peaks are much reduced with respect to those of the LT-GaAs. This narrowing may be brought about by a reduction of the electrostatic disorder due to an increasingly stronger competition of Mn with As for Ga sites.[35]

For the (Ga,Mn)As layers with $x \geq 0.3\%$ the structure of the LT-PR spectra is significantly changed, only two features are seen. In the $x = 0.3\%$ layer the lower-energy peak is red shifted with respect to the heavy hole feature in the spectrum of $x = 0.0007\%$ layer. However, for the other layers the low energy peak, which becomes dominant, slightly shifts to higher energies and the separation between the two features increases with $x$. Moreover, the two features become broaden and their relative intensities indicate modification of the VB edge with increasing Mn content as well.

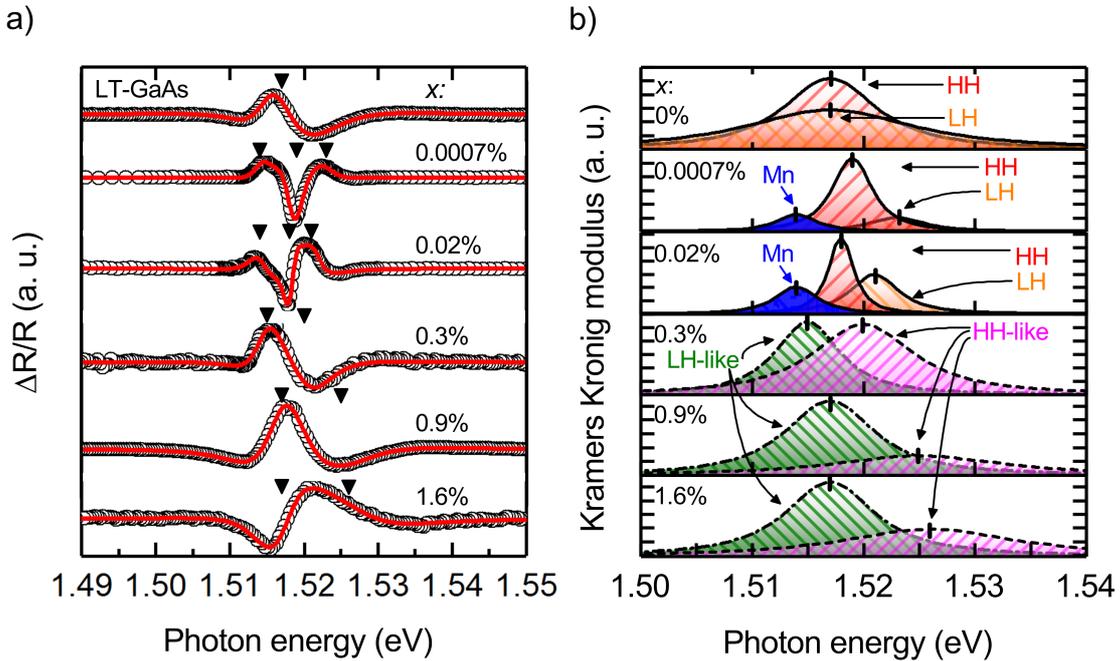

**Figure 5.** (a) Normalized low-temperature photoreflectance spectra $\Delta R/R$ of the LT-GaAs and (Ga,Mn)As layers (open circles). Solid line indicates cumulative fit of number of fitted third derivative line-shape (TDLS) features. The transition energies are denoted by arrows. Parameters of the single TDLS components of each spectrum have been used to calculate Kramers-Kronig (KK) modulus. (b) Results of KK integration of the LT-PR spectra fitted components. For $x = 0\%$ (LT-GaAs layer) the two, heavy-hole (HH) and light-hole (LH) sub-bands related, features can be observed (diagonally hatched). An addition of Mn to GaAs changes the spectra by appearance of the low energy feature (solid filled one, marked with "Mn") together with the energy separation of the rest of the features. Turning the conductivity, from $n$ to $p$-type, for $x \geq 0.3\%$, modifies the character of the spectra into two broadened features instead of three. Generally, with increasing $x$ the spectra tend to move to higher energies.

A complementary information about the (Ga,Mn)As band structure can be obtained from angle-resolved photoemission measurements. As long as the photoreflectance spectroscopy probes bulk properties of (Ga,Mn)As layers and the surface conditions do not significantly influence the spectra, the photoemission experiment with low photon energies (below 100 eV, as applied here) is surface sensitive. Earlier studies of the VB dispersion were conducted via *ex-situ* ARPES.[18,19,21] Recently, ARPES (using a portable vacuum transfer chamber) and He discharge lamp as the incident radiation source was used to study the Fermi level position in (Ga,Mn)As by Souma *et al.*[57] Similarly, only very limited number of integrated photoemission studies on synchrotron beamlines were performed using the samples transferred from MBE growth chambers without air exposure.[20] Because of the significant influence of the growth conditions[58] and surface quality on the experimental spectra the authors of measurements using synchrotron radiation came to contradicting conclusions.[18,19,57,59–61] It has to be stressed that due to the high reactivity of (Ga,Mn)As surface and the metastable character of this compound the only reliable, ARPES measurements are those which have been done either with use of the MBE growth setup vacuum connected to ARPES analytical chamber, or with the samples transferred between MBE and ARPES in ultrahigh vacuum conditions (e.g. in vacuum suitcase), on the condition that this is done properly. It has been demonstrated repeatedly (ref. 60 and the references therein), that ARPES results obtained with (Ga,Mn)As samples exposed to air, or even protected with amorphous As capping, and subjected to thermal annealing and/or $Ar^+$ ion bombardment to get clean oxide-free surface indispensable for ARPES measurements (e.g. ref. 18, 59) do contain artifacts hampering correct interpretation of the ARPES data (ref. 60 and discussion therein).

To avoid the issues of the metastable character of (Ga,Mn)As solid solution (it decomposes when annealed at temperatures only slightly exceeding the growth temperatures[58,62]) and of a high reactivity of Mn-rich (Ga,Mn)As surface,[60,63] we present here results of the *in-situ* ARPES measurements. The measurements have been done at I3 beamline at MAX IV national Swedish synchrotron radiation facility, with the MBE growth setup directly connected to the ARPES system. To this end two freshly grown samples: the $x = 1.6\%$ ferromagnetic (Ga,Mn)As and the reference LT-GaAs layers have been transferred directly (i.e. in an ultrahigh vacuum environment) from the SVTA III-V MBE growth chamber to the ARPES setup.[60] The region of the Fermi level, $E_{Fermi}$, and the VB maximum, $E_{max}$, have been analysed at about 20 K and the results are displayed in Fig. 6.

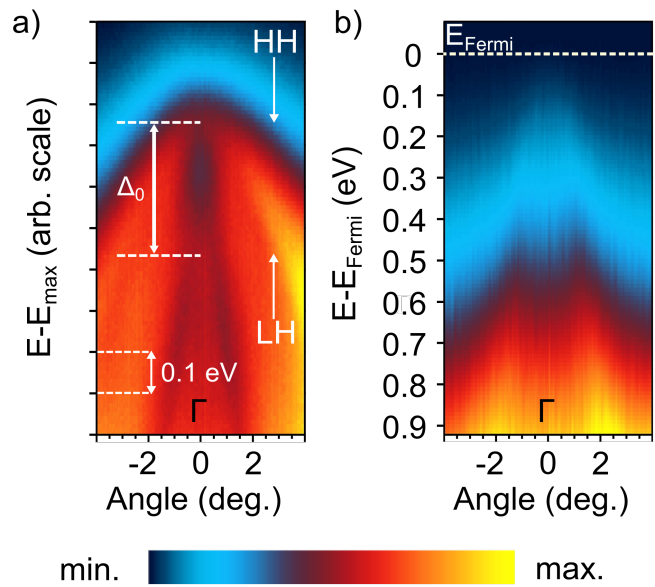

**Figure 6.** Low temperature (≈20 K) ARPES of a) LT-GaAs and b) (Ga,Mn)As 1.6% Mn. Heavy (HH) and light (LH) hole sub-bands as well as split-off energy ($\Delta_0$) are denoted by arrows in a) - distinguishable only in the case of LT-GaAs spectrum. Low-doped ferromagnetic (Ga,Mn)As epitaxial layer spectrum reveals drop of the photoemission intensity below the Fermi level what does not allow to distinguish GaAs VB features.

In the case of LT-GaAs the spectrum clearly reflects the curvature of the VB dispersion with heavy hole and split-off bands. Fermi level is located within the band gap due to an amount of As$_{Ga}$ donor defects originating from low temperature growth conditions.[60] In Fig. 6b it is clearly visible that Mn doping significantly modifies the VB structure of (Ga,Mn)As below the Fermi level and causes a deformation of the spectrum what does not allow to establish clear band edge of split-off band as in the case of LT-GaAs (Fig. 6a). The lack of the sharp band edges (as compared to LT-GaAs) and systematic decrease in the photoemission signal with approaching $E_{Fermi}$ can be observed. Moreover, there is no increase in the photoemission signal near the Fermi level, which precludes the existence of a detached band -related density of states. The ARPES spectra presented here are in agreement with those of ref. 60 and recent results from ref. 57 for low $T_C$ (Ga,Mn)As and thus low influence of the Coulomb blockade on the density of states near the Fermi level can be inferred.

**Discussion and conclusions**

All the experimental findings presented here converge into a cohesive picture of the evolution of the electronic properties of (Ga,Mn)As in the low concentration limit. The studied concentration range, from $x$ = 0% to 1.6% proved sufficient to span both the "intrinsic" – presumably *n*-type and the *p*-type conductivity limits and to identify Mn concentration 0.02% < $x_b$ < 0.3%, to be the bordering one. Most of the features observed in the layers with $x < x_b$ are very similar (if not identical) to those observed in the reference LT-GaAs layer. This is the case of the HR-XRD, the micro-Raman spectroscopy, and the sensitive SQUID magnetometry. However, the Mn-related impurity states have been revealed by the LT-PR studies, which, in addition to two higher-energy HH and LH peaks, yielded an additional below-band-gap peak. The presence of Mn extended states in the band gap influences the hybridization of the VB states what leads to the HH – LH splitting observed as a low and high energy peaks in the KK analysis of LT-PR. The comparison between the LT-GaAs and $x < x_b$ (Ga,Mn)As case are schematically sketched in Figs. 7a and 7b, respectively.

The LT-PR picture changes completely when $x$ exceeds $x_b$. Now, only two broadened features are seen. The lack of the optical transition related to filled Mn impurity states and a broadening of the two remaining optical transitions are indicative that Fermi level has nested in the VB. Indeed, as established by HR-XRD, the increasing incorporation of Mn compressively strains the (Ga,Mn)As layers. Such relatively small changes of the resulting out-of-plane strain (Tab. 1) indicate its rather small influence on the energy positions of the spectral features in the LT-PR spectra. Considering the shear potential (b = -2 eV), elastic constants ($C_{11}$ = 119 GPa, $C_{12}$ = 53.8 GPa) and spin-orbit splitting ($\Delta_0$ = 0.34 eV) for GaAs the HH-LH splitting at $k$ = 0 can be estimated to be approximately 2 meV even for $x$ = 1.6% - much smaller than we observe. Therefore, the splitting of the two features can be explained by the transitions occurring at $k \neq 0$ making HH-CB transition higher in energy than LH-CB due to Fermi level position, as sketched in Fig. 7c. In the $x$ = 0.3% case the observed red shift is due to the carrier-induced many-body band-gap renormalization (BGR),[28,53,64] which prevails over the band filling for such a low *p*. Accordingly, the similar magnitudes of the peaks in the spectra support a shallow Fermi level position in the VB. In line with such an interpretation the low and high energy peaks in the KK analysis can be attributed to LH-CB and HH-CB optical transitions, respectively. Such a scenario calls for the *p*-type character of all $x \geq 0.3\%$ layers and indeed it is corroborated by the micro-Raman studies, which indicated the presence of the hole-related CPPM structure whose intensity increases with $x$. However, on increasing $x$, an expected sizable blue shift of the optical gap due to the Moss-Burstein effect has not been observed. This is a further consequence of the BGR- related narrowing effects which also arise with the increase in $p$. Nevertheless, the lowering of the Fermi level postulated here is directly reflected in the changes of the amplitude ratio and the broadening of these two LT-PR peaks, the effects expected for an increasing magnitude of the $k$ vector for higher energy transitions. It is due to the fact that the optical transition probability decreases with $k$ vector value moving further away from the critical point of the band structure ($k = 0$, $\Gamma$ point of the Brillouin zone). The picture of the band structure drawn by this interpretation is supported by the *in-situ* ARPES studies which revealed perturbed but unitary structure below the Fermi level for $x = 1.6\%$ ferromagnetic sample.

In parallel, the increase in the Mn incorporation into GaAs results in a remarkable evolution of the magnetic properties from a weak paramagnet through the superparamagnetic-like inhomogeneous ferromagnet (with the coupling being effective on 10-20 nm length scale), up to the long range (global) itinerant ferromagnetic material. The experience accumulated so far allows saying that these three basic types of magnetism can coexist in any sample in a degree depending on the Mn concentration and its spatial distribution, growth protocol, thickness of the layer, its post growth thermal history, and on the strength of various physical fields applied – to name the most significant parameters. The research presented here points clearly out that there is a kind of a threshold $p/x$ ratio, or a maximum compensation degree above which neither the global nor the mesoscopic FM coupling develops in DFS. This is seen in our $x = 0.3\%$ layer. The Raman studies prove that it is $p$-type, and it owes this to a rather uncommonly met very low concentration of hole-compensating donor $As_{Ga}$ defects, as indicated by HR-XRD. However, its hole density is so low that it does not exhibit any signs of a FM coupling above 2 K. It remains paramagnetic as long as magnetic studies at $T \geq 2$ K can tell. This fact implies that the Mn-induced modification of the (Ga,Mn)As VB observed in the LT-PR occurs for Mn content much lower than that, required for the development of the ferromagnetic coupling, as predicted by recent calculations.[65] On the other hand, and perhaps most importantly, the combined magnetic and LT-PR studies presented here univocally indicate that the PM ↔ FM transformation in carrier mediated DFS takes place without corresponding or underlying changes of the character of the valence (in the case of holes) band. The whole process is rooted in the nanoscale fluctuations of the local density of states (LDOS) and the SP-like "phase" can be regarded as the cornerstone of the DFS magnetism, serving as the pillar supporting the bridge spanning both PM and FM thermodynamic phases. It is worth noting that the nanometre extent of the FM coupling of the SP component, already seen before in ref. 2, 46, 66, 67, and confirmed here, agrees well with the extent of the spatial variations in the LDOS in the VB recognized in a similar (Ga,Mn)As layer by a scanning tunnelling microscopy.[68] The mesoscopic character of this magnetism calls rather for a proper quantum mechanical description.

Another point worth strengthening here is the role of the correct approach of the magnetic measurements. It is relatively easy to assign the signal of the SP-like response to the FM one. The most prominent example comes from Fig. 3 which indicates that if the "zero field" measurements were performed at 30 Oe, that is if the "remnant moment" was probed at 30, or even 50 Oe - as some authors did,[69,70] the SP nature of this $p = 5 \times 10^{19}$ cm$^{-3}$ sample would have never been unveiled, moreover, this sample would have been classified as a truly FM one with a "$T_C$" of about 12 - 15 K, falsely directing to wrong conclusions.

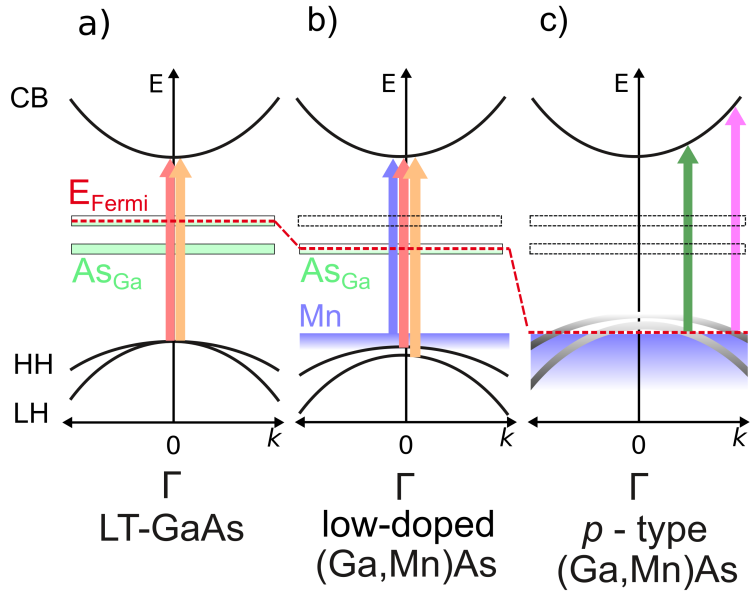

**Figure 7.** Schematic valence-band diagram for a) LT-GaAs, b) *n*-type and c) *p*-type (Ga,Mn)As. Arrows indicate inter-band optical transitions. For LT-GaAs the transitions originating from heavy- and light-hole sub-bands (denoted by HH and LH, respectively) to conduction band (CB) at the Γ critical point of the Brillouin zone are indicated. For *n*-type (Ga,Mn)As an additional below-band-gap feature, arising from the filled Mn states, appears together with a small splitting of the valence sub-bands, and the Fermi level moves downward with increasing Mn content due to the compensation of As$_{Ga}$-related donors. For *p*-type (Ga,Mn)As the Fermi level lies within the strongly disturbed VB, which, in turn, is extended into the band gap as a result of the carrier-induced many-body effects. The Fermi level position in *p*-type (Ga,Mn)As is determined by the valence-band filling and many-body effects.

In brief, our results of the optical-energy-gap measurements supported by the complementary characterization provide an important insight into the dispute on the Fermi level behaviour in (Ga,Mn)As. By applying the high-spectral-resolution spectroscopic studies, supplemented with electronic, magnetic and structural characterization, we have found that modification of the GaAs VB caused by Mn incorporation occurs already for a very low Mn content, much lower than that required to support long range ferromagnetic spin – spin coupling in (Ga,Mn)As. Mn doping creates itinerant hole plasma within the structurally unitary VB. Therefore the magnetism observed in (Ga,Mn)As is driven by the VB holes however its fine details are strongly dependent on the local electronic configuration. Only for *n*-type (Ga,Mn)As the Mn-related extended states are visible as a feature detached from the valence-band edge and partly occupied with electrons.

**Methods**

100 nm thick (Ga,Mn)As and LT-GaAs epitaxial layers have been grown using low temperature MBE at approximately 230°C on GaAs (001) semi-insulating substrates. Tailoring the substrate temperature depending on the intentional Mn content has enabled to maximize $Mn_{sub}$ incorporation into GaAs matrix and reduce $As_{Ga}$ and $Mn_I$ defects concentration. Additionally, for photoemission studies LT-GaAs and (Ga,Mn)As layer with $x$ = 1.6% have been also grown on $p$-type conductive substrates to prevent charging effects. The MBE growth has been done with $As_2$ flux, at optimized conditions, i.e. with $As_2$/(Ga+Mn) flux ratio close to the stoichiometric one, as carefully set during the preceding growth of test/calibration samples. The high perfection of a 2-D layer-by-layer growth of (Ga,Mn)As has been confirmed by RHEED intensity oscillations, usually observed for at least first 50 nm of the grown layer (in some cases up to the very end of the growth of 100 nm thick films).[71] Mn contents in (Ga,Mn)As were confirmed by SIMS and HR-XRD of Cu K$\alpha_1$.

Micro-Raman spectroscopy experiment was done using 514 nm Ar$^+$ ion laser line in backscattering configuration at room temperature with the resolution of 1 cm$^{-1}$.

The magnetic properties were investigated using a Quantum Design MPMS XL SQUID magnetometer equipped with a low field option. The measurements were carried out down to 1.8 K, the base temperature of the SQUID system, and up to 5 kOe. Important for this study the truly near-zero field conditions in the magnetometer ($H \cong 0.1$ Oe, as established using a $Dy_2O_3$ paramagnetic salt) were achieved in two steps. Firstly, before the low field measurements the magnetometer was degaussed with an oscillating magnetic field of decreasing amplitude. Secondly, a soft quench of the SQUID's superconducting magnet was routinely performed prior to the zero-field studies such as the thermoremnant moment (TRM, the measurement of the remnant moment on increasing $T$) and during thermal cycling of the sample brought beforehand to its remanence. Each such cycle consisted of warming up of the sample to a progressively higher temperature followed by re-cooling to 1.8 K, following the pattern sketched in the inset in Fig. 3. The latter measurement allows to distinguish a decaying (with temperature) part of the sample remnant moment [that is the dynamically blocked one by energy barriers] from that related to the spontaneous magnetization in the equilibrium state under the zero field conditions, if of course such contributions exist in the specimen. All these magnetic measurements were carried out using about ~20 cm long silicon strips to fix the samples in the magnetometer and the adequate experimental code for minute signals measurements was strictly observed.[43]

Photoreflectace spectroscopy measurements were performed using 371 nm pump beam wavelength ensuring that modulation signal originates from the top part of investigated layers. Spectrometer was set-up into so the called "dark" configuration with a halogen lamp coupled with a monochromator as a probe beam source. The signal was detected by a silicon photodiode. The sample was cooled with an optical closed-cycled helium cryostat.

The ARPES was measured at MAX-IV synchrotron I3 beam line vacuum-connected to the SVTA III-V MBE chamber. The sample was examined also by the low energy electron diffraction in ARPES preparation chamber which shown c(4×4) and (1×2) surface reconstruction patterns for LT-GaAs and (Ga,Mn)As, respectively. For photoemission experiments we have used the excitation energy of 22.5 eV and p-polarization. For ARPES measurement, the samples were cooled-down to about 20 K. For (Ga,Mn)As spectra we used smoothing procedure based on cubic spline function fitting. Fermi level has been determined using metallic Mn stripe located at the boarder of the wafer.

Acknowledgements

This study has been supported by the Foundation for Polish Science under Grant No. POMOST/2010-2/12 sponsored by the European Regional Development Fund, National Cohesion Strategy: Innovative Economy; National Science Centre (Poland) through MAESTRO Grant No. 2011/02/A/ST3/00125 and OPUS Grant No. 2014/13/B/ST3/04489. ARPES and MBE projects acknowledge the support from the Baltic Science Link coordinated by the Swedish Research Council (VR). We would like to thank Mats Leandersson and Balasubramanian Thiagarajan for a support in ARPES measurements at I3 beamline of MAX-IV facility and T. Dietl for critical reading of the manuscript.

Author Contributions statement:

L.G., O.Y. and J.Z. collected and analysed LT-PR and Raman data. J.S. grown the samples. M.S. and T.A. performed SQUID measurements, while M.S. provided with their interpretation. O.Y., L.G. and J.S. measured ARPES. JZ.D carried out HR-XRD measurements. R.J. performed SIMS measurements. L.G., O.Y., T.W., M.S., J.Z. and J.S. wrote the manuscript. All co-authors contributed to the discussion, revision of the manuscript and reply Reviewers' comments.

1. Dietl, T. & Ohno, H. Dilute ferromagnetic semiconductors: Physics and spintronic structures. *Rev. Mod. Phys.* **86,** 187-251 (2014).

2. Sawicki, M. *et al.* Experimental probing of the interplay between ferromagnetism and localization in (Ga,Mn)As. *Nat. Phys* **6,** 22-25 (2010).

3. Ohno, H. *et al.* Electric-field control of ferromagnetism. *Nature* **408,** 944-946 (2000).

4. Stolichnov, I. *et al.* Non-volatile ferroelectric control of ferromagnetism in (Ga,Mn)As. *Nat. Mater.* **7,** 464-467 (2008).

5. Chiba, D., Ono, T., Matsukura, F. & Ohno, H. Electric field control of thermal stability and magnetization switching in (Ga,Mn)As. *Appl. Phys. Lett.* **103,** 142418 (2013).

6. Oiwa, A., Słupinski, T. & Munekata, H. Control of magnetization reversal process by light illumination in ferromagnetic semiconductor heterostructure p-(In,Mn)As/GaSb. *Appl. Phys. Lett.* **78,** 518-520 (2001).

7. Li, Z. *et al.* Pulsed field induced magnetization switching in (Ga,Mn)As. *Appl. Phys. Lett.* **92,** 112513 (2008).

8. Rozkotová, E. *et al.* Light-induced magnetization precession in GaMnAs. *Appl. Phys. Lett.* **92,** 122507 (2008).

9. Gryglas-Borysiewicz, M. *et al.* Hydrostatic pressure study of the paramagnetic-ferromagnetic phase transition in (Ga,Mn)As. *Phys. Rev. B* **82,** 153204 (2010).

10. Gryglas-Borysiewicz, M. *et al.* Hydrostatic-pressure-induced changes of magnetic anisotropy in (Ga, Mn)As thin films. *J. Phys. Condens. Matter* **29,** 115805 (2017).

11. Wolf, S. A. Spintronics: A Spin-Based Electronics Vision for the Future. *Science* **294,** 1488-1495 (2001).

12. Dietl, T. Zener Model Description of Ferromagnetism in Zinc-Blende Magnetic Semiconductors. *Science* **287,** 1019-1022 (2000).

13. Dietl, T. A ten-year perspective on dilute magnetic semiconductors and oxides. *Nat. Mater.* **9,** 965-974 (2010).

14. Alberi, K. *et al.* Formation of Mn-derived impurity band in III-Mn-V alloys by valence band anticrossing. *Phys. Rev. B* **78,** 075201 (2008).

15. Ohya, S., Takata, K. & Tanaka, M. Nearly non-magnetic valence band of the ferromagnetic semiconductor GaMnAs. *Nat. Phys.* **7,** 342-347 (2011).

16. Ando, K., Saito, H., Agarwal, K. C., Debnath, M. C. & Zayets, V. Origin of the Anomalous Magnetic Circular Dichroism Spectral Shape in Ferromagnetic $Ga_{1-x}Mn_xAs$ : Impurity Bands inside the Band Gap. *Phys. Rev. Lett.* **100,** 067204 (2008).

17. Muneta, I., Terada, H., Ohya, S. & Tanaka, M. Anomalous Fermi level behavior in GaMnAs at the onset of ferromagnetism. *Appl. Phys. Lett.* **103,** 032411 (2013).

18. Kobayashi, M. *et al.* Unveiling the impurity band induced ferromagnetism in the magnetic semiconductor (Ga,Mn)As. *Phys. Rev. B* **89,** 205204 (2014).

19. Gray, A. X. *et al.* Bulk electronic structure of the dilute magnetic semiconductor $Ga_{1-x}Mn_xAs$ through hard X-ray angle-resolved photoemission. *Nat. Mater.* **11,** 957–962 (2012).

20. Di Marco, I. *et al.* Electron correlations in $Mn_xGa_{1-x}As$ as seen by resonant electron spectroscopy and dynamical mean field theory. *Nat. Commun.* **4,** 2645 (2013).

21. Fujii, J. *et al.* Identifying the Electronic Character and Role of the Mn States in the Valence Band of (Ga,Mn)As. *Phys. Rev. Lett.* **111,** 097201 (2013).

22. Chapler, B. C. *et al.* Ferromagnetism and infrared electrodynamics of $Ga_{1-x}Mn_xAs$. *Phys. Rev. B* **87,** 205314 (2013).

23. Chapler, B. C. et al. Infrared probe of the insulator-to-metal transition in $Ga_{1-x}Mn_xAs$ and $Ga_{1-x}Be_xAs$. *Phys. Rev. B* **84,** 081203 (2011).

24. Muneta, I., Ohya, S., Terada, H. & Tanaka, M. Sudden restoration of the band ordering associated with the ferromagnetic phase transition in a semiconductor. *Nat. Commun.* **7,** 12013 (2016).

25. Pollak, F. H. & Shen, H. Modulation spectroscopy of semiconductors: bulk/thin film, microstructures, surfaces/interfaces and devices. *Mater. Sci. Eng. R Reports* **10,** 275-374 (1993).

26. Yastrubchak, O. et al. Effect of low-temperature annealing on the electronic- and band-structures of (Ga,Mn)As epitaxial layers. *J. Appl. Phys.* **115,** 012009 (2014).

27. Yastrubchak, O. et al. Photoreflectance study of the fundamental optical properties of (Ga,Mn)As epitaxial films. *Phys. Rev. B* **83,** 245201 (2011).

28. Yastrubchak, O. et al. Electronic- and band-structure evolution in low-doped (Ga,Mn)As. *J. Appl. Phys.* **114,** 053710 (2013).

29. Yastrubchak, O. et al. Ferromagnetism and the electronic band structure in (Ga,Mn)(Bi,As) epitaxial layers. *Appl. Phys. Lett.* **105,** 072402 (2014).

30. de Boer, T., Gamouras, A., March, S., Novák, V. & Hall, K. C. Observation of a blue shift in the optical response at the fundamental band gap in $Ga_{1-x}Mn_xAs$. *Phys. Rev. B* **85,** 033202 (2012).

31. Szczytko, J., Mac, W., Twardowski, A., Matsukura, F. & Ohno, H. Antiferromagnetic p−d exchange in ferromagnetic $Ga_{1-x}Mn_xAs$ epilayers. *Phys. Rev. B* **59,** 12935 (1999).

32. Tesařová, N. et al. Systematic study of magnetic linear dichroism and birefringence in (Ga,Mn)As. *Phys. Rev. B* **89,** 085203 (2014).

33. Yildirim, M. et al. Electronic structure of $Ga_{1-x}Mn_xAs$ probed by four-wave mixing spectroscopy. *Phys. Rev. B* **84,** 121202(R) (2011).

34. Burch, K. S., Stephens, J., Kawakami, R. K., Awschalom, D. D. & Basov, D. N. Ellipsometric study of the electronic structure of $Ga_{1-x}Mn_xAs$ and low-temperature GaAs. *Phys. Rev. B* **70,** 205208 (2004).

35. Sadowski, J. & Domagala, J. Z. Influence of defects on the lattice constant of GaMnAs. *Phys. Rev. B* **69,** 075206 (2004).

36. Yastrubchak, O. et al. Misfit strain anisotropy in partially relaxed lattice-mismatched InGaAs/GaAs heterostructures. *J. Phys. Condens. Matter* **16,** S1-S8 (2004).

37. Yastrubchak, O. et al. Strain release in $InGaAs/In_xAl_{1-x}As/InP$ heterostructures. *Phys. B Condens. Matter* **340,** 1082-1085 (2003).

38. Limmer, W. *et al.* Optical Study of Plasmon–LO Phonon Modes in $Ga_{1-x}Mn_xAs$. *J. Supercond.* **17,** 417-420 (2004).

39. Look, D. C. On compensation and conductivity models for molecular-beam-epitaxial GaAs grown at low temperature. *J. Appl. Phys.* **70,** 3148-3151 (1991).

40. Wenzel, M., Irmer, G., Monecke, J. & Siegel, W. Determination of the effective Hall factor in p-type semiconductors. *Semicond. Sci. Technol.* **13,** 505-511 (1998).

41. Irmer, G., Wenzel, M. & Monecke, J. Light scattering by a multicomponent plasma coupled with longitudinal-optical phonons: Raman spectra of p-type GaAs:Zn. *Phys. Rev. B* **56,** 9524-9538 (1997).

42. Jungwirth, T. *et al.* Character of states near the Fermi level in (Ga,Mn)As: Impurity to valence band crossover. *Phys. Rev. B* **76,** 125206 (2007).

43. Sawicki, M., Stefanowicz, W. & Ney, A. Sensitive SQUID magnetometry for studying nanomagnetism. *Semicond. Sci. Technol.* **26,** 064006 (2011).

44. Ney, A., Harris, J. S. & Parkin, S. S. P. Temperature dependent magnetic properties of the GaAs substrate of spin-LEDs. *J. Phys. Condens. Matter* **18,** 4397-4406 (2006).

45. Sawicki, M. Magnetic properties of (Ga,Mn)As. *J. Magn. Magn. Mater.* **300,** 1-6 (2006).

46. Šiušys, A. *et al.* All-Wurtzite (In,Ga)As-(Ga,Mn)As Core–Shell Nanowires Grown by Molecular Beam Epitaxy. *Nano Lett.* **14,** 4263-4272 (2014).

47. Dietl, T. Interplay between Carrier Localization and Magnetism in Diluted Magnetic and Ferromagnetic Semiconductors. *J. Phys. Soc. Japan* **77,** 031005 (2008).

48. Hrabovsky, D. *et al.* Magnetization reversal in GaMnAs layers studied by Kerr effect. *Appl. Phys. Lett.* **81,** 2806-2808 (2002).

49. Sawicki, M. *et al.* In-plane uniaxial anisotropy rotations in (Ga,Mn)As thin films. *Phys. Rev. B* **71,** 121302 (2005).

50. Stefanowicz, W. *et al.* Magnetic anisotropy of epitaxial (Ga,Mn)As on (113) A GaAs. *Phys. Rev. B* **81,** 155203 (2010).

51. Jungwirth, T. *et al.* Prospects for high temperature ferromagnetism in (Ga,Mn)As semiconductors. *Phys. Rev. B* **72,** 165204 (2005).

52. Wang, K. Y. *et al.* Influence of the Mn interstitial on the magnetic and transport properties of (Ga,Mn)As. *J. Appl. Phys.* **95,** 6512-6514 (2004).

53. Dietl, T., Ohno, H. & Matsukura, F. Hole-mediated ferromagnetism in tetrahedrally coordinated semiconductors. *Phys. Rev. B* **63,** 195205 (2001).

54. Aspnes, D. E. Third-derivative modulation spectroscopy with low-field electroreflectance. *Surf. Sci.* **37,** 418-442 (1973).

55. Gluba, L. *et al.* On the nature of the Mn-related states in the band structure of (Ga,Mn)As alloys via probing the $E_1$ and $E_1 + \Delta_1$ optical transitions. *Appl. Phys. Lett.* **105,** 032408 (2014).

56. Jezierski, K. *et al.* Application of Kramers–Krönig analysis to the photoreflectance spectra of heavily doped GaAs/SI-GaAs structures. *J. Appl. Phys.* **77,** 4139-4141 (1995).

57. Souma, S. *et al.* Fermi level position, Coulomb gap, and Dresselhaus splitting in (Ga,Mn)As. *Sci. Rep.* **6,** 27266 (2016).

58. Sadowski, J. *et al.* Formation process and superparamagnetic properties of (Mn,Ga)As nanocrystals in GaAs fabricated by annealing of (Ga,Mn)As layers with low Mn content. *Phys. Rev. B* **84,** 245306 (2011).

59. Okabayashi, J. *et al.* Angle-resolved photoemission study of $Ga_{1-x}Mn_xAs$. *Phys. Rev. B* **64,** 125304 (2001).

60. Kanski, J. *et al.* Electronic structure of (Ga,Mn)As revisited. *New J. Phys.* **19,** 023006 (2017).

61. Souma, S. *et al.* Comment on 'Alternative interpretation of the recent experimental results of angle-resolved photoemission spectroscopy on GaMnAs [Sci. Rep. **6,** 27266 (2016)]' by M. Kobayashi et al., arXiv:1608.07718.

62. Jeong, Y. *et al.* Effect of thermal annealing on the magnetic anisotropy of GaMnAs ferromagnetic semiconductor. *Curr. Appl. Phys.* **14,** 1775-1778 (2014).

63. Bartoš, I. *et al.* Mn incorporation into the GaAs lattice investigated by hard x-ray photoelectron spectroscopy and diffraction. *Phys. Rev. B* **83,** 235327 (2011).

64. Zhang, Y. & Das Sarma, S. Temperature and magnetization-dependent band-gap renormalization and optical many-body effects in diluted magnetic semiconductors. *Phys. Rev. B* **72,** 125303 (2005).

65. Milowska, K. Z. & Wierzbowska, M. Hole $sp^3$-character and delocalization in (Ga,Mn)As revised with pSIC and MLWF approaches – Newly found spin-unpolarized gap states of s-type below 1% of Mn. *Chem. Phys.* **430,** 7-12 (2014).

66. Schlapps, M. *et al.* Transport through (Ga, Mn) As nanoislands: Coulomb blockade and temperature dependence of the conductance. *Phys. Rev. B* **80,** 125330 (2009).

67. Sadowski, J. *et al.* Wurtzite (Ga, Mn) As nanowire shells with ferromagnetic properties. *Nanoscale* **9,** 2129-2137 (2017).

68. Richardella, A. *et al.* Visualizing Critical Correlations Near the Metal-Insulator Transition in $Ga_{1-x}Mn_xAs$. *Science* **327,** 665-669 (2010).

69. Sheu, B. L. *et al.* Onset of Ferromagnetism in Low-Doped $Ga_{1-x}Mn_xAs$. *Phys. Rev. Lett.* **99,** 227205 (2007).

70. Mayer, M. A. *et al.* Electronic structure of $Ga_{1-x}Mn_xAs$ analyzed according to hole-concentration-dependent measurements. *Phys. Rev. B* **81,** 045205 (2010).

71. Sadowski, J. *et al.* Structural and magnetic properties of molecular beam epitaxy grown GaMnAs layers. *J. Vac. Sci. Technol. B Microelectron. Nanom. Struct.* **18,** 1697-1700 (2000).

.

On the origin of magnetism in (Ga,Mn)As: from paramagnetic through superparamagnetic to ferromagnetic phase

L. Gluba[1,2], O. Yastrubchak[3,2*], J.Z. Domagala[4], R. Jakiela[4], T. Andrearczyk[4], J. Żuk[2], T. Wosinski[4], J. Sadowski[5,4,6] and M. Sawicki[4]

[1]Institute of Agrophysics, Polish Academy of Sciences, Doświadczalna 4, 20-290 Lublin, Poland

[2]Institute of Physics, Maria Curie-Sklodowska University in Lublin, Pl. M. Curie-Skłodowskiej 1, 20-031 Lublin, Poland

[3]V.E. Lashkaryov Institute of Semiconductor Physics, National Academy of Sciences of Ukraine, 03028 Kyiv, Ukraine

[4]Institute of Physics, Polish Academy of Sciences, Aleja Lotnikow 32/46, PL-02668 Warsaw, Poland

[5]MAX-IV laboratory, Lund University, P.O. Box.118, 22100 Lund, Sweden

[6]Department of Physics and Electrical Engineering, Linnaeus University, SE-391 82 Kalmar, Sweden

The single resonance feature of a photoreflectance spectra, in the low-field regime, can be expressed as:[1]

$$\frac{\Delta R}{R} = \Re\left[ Ce^{i\theta}(\hbar\omega - E_{CP} + i\Gamma)^{-n} \right], \qquad (1)$$

where $C$ is an amplitude parameter, $\theta$ is the phase factor determining asymmetry of the line-shape, $\hbar\omega$ is the photon energy, $E_{CP}$ is the optical transition energy (critical point), $\Gamma$ is the width of the structure. Index $n$ determine dimensionality of the critical point. For three dimensional one $n = 5/2$. The expression (1) is called third derivative line-shape (TDLS). Fig. S1 shows the results of fitting TDLS features to the $\Delta R/R$ photoreflectance spectra of LT-GaAs and (Ga,Mn)As layers. The deconvolution results have been used to convert derivative like spectral features into the peaks for interpretation clarity. It has been done by calculating Kramers-Kronig integral of $\Delta R/R$ spectra, $\Delta\rho_I$, defined as:[2]

$$\Delta\rho_I(E_{CP}) = \frac{2E_{CP}}{\pi} \int_{E_b}^{E_a} \frac{\Delta R}{R} \frac{1}{E_{CP}^2 + E^2} dE, \qquad (2)$$

---

* Corresponding author: Oksana Yastrubchak; E-mail: plazmonoki@gmail.com

where $E_a$ and $E_b$ are boundaries of the ΔR/R spectra and $ΔR/R(E_a) = ΔR/R(E_b) = 0$. Finally, using $Δρ_I(E_{CP})$, we have calculated Kramers-Kronig modulus, $Δρ$:[2]

$$\Delta \rho = \sqrt{(\Delta R/R)^2 + (\Delta \rho_I)^2}, \qquad (3)$$

which has been shown in Fig. 5b of the main text of the article.

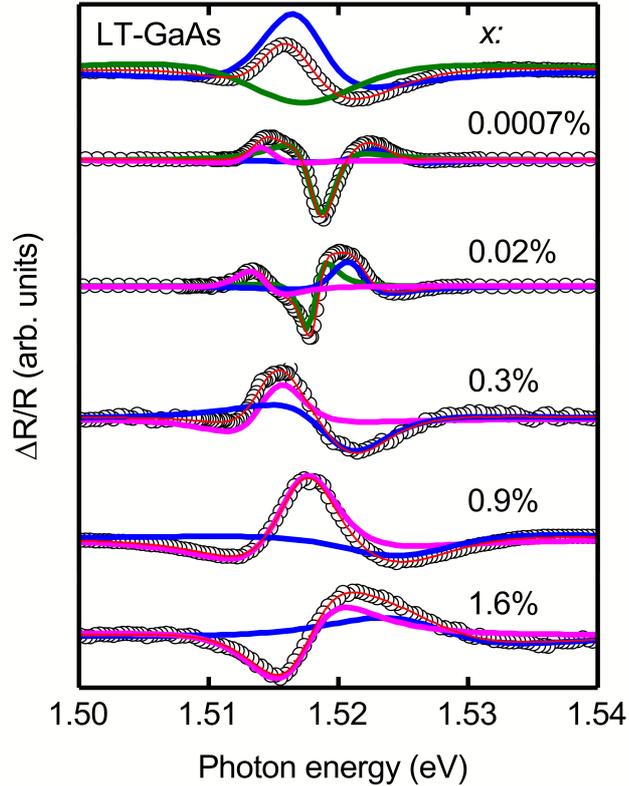

Figure S1. Deconvolution of the low-temperature photoreflectance spectra ΔR/R of the LT-GaAs and (Ga,Mn)As layers (open circles). Solid lines indicate single third derivative line-shape (TDLS) features contributing to the cumulative fitting line. Above depicted TDLS features have been used to calculate Kramers-Kronig modulus shown in Fig. 5 b).

Aspnes, D. E. Third-derivative modulation spectroscopy with low-field electroreflectance. *Surf. Sci.* **37,** 418-442 (1973).

*Jezierski, K. et al. Application of Kramers–Krönig analysis to the photoreflectance spectra of heavily doped GaAs/SI-GaAs structures. J. Appl. Phys.* **77,** *4139-4141 (1995).*